\documentclass[11pt,a4paper]{article}
\pdfoutput=1
\usepackage{jheppub}
\usepackage{amsmath,amssymb,mathtools}

\usepackage{xcolor}

\newcommand{\SO}{{\rm SO}}

\usepackage{tensor}

\title{Higher-Derivative Supergravity, Wrapped M5-branes, and Theories of Class $\mathcal{R}$}
\author[{\rm A}]{Nikolay Bobev,}
\author[{\rm A}]{Anthony M. Charles,}
\author[{\rm D}]{Dongmin Gang,}
\author[{\rm E}]{Kiril Hristov,}
\author[{\rm A}]{and Valentin Reys}

\affiliation[{\rm A}]{Institute for Theoretical Physics, KU Leuven, \\
Celestijnenlaan 200D, B-3001 Leuven, Belgium}

\affiliation[{\rm D}]{Asia Pacific Center for Theoretical Physics (APCTP),\\ 
Pohang 37673, Korea}

\affiliation[{\rm D}]{ Department of Physics, Pohang University of Science and Technology (POSTECH), \\
Pohang 37673, Korea}

\affiliation[{\rm E}]{INRNE, Bulgarian Academy of Sciences,\\
Tsarigradsko Chaussee 72, 1784 Sofia, Bulgaria}

\affiliation[{\rm E}]{Faculty of Physics, Sofia University,\\
J. Bourchier Blvd. 5, 1164 Sofia, Bulgaria}

\emailAdd{nikolay.bobev, anthony.charles, valentin.reys @kuleuven.be}
\emailAdd{dongmin.gang@apctp.org}
\emailAdd{khristov@phys.uni-sofia.bg}

\abstract{We study the interplay between four-derivative 4d gauged supergravity, holography, wrapped M5-branes, and theories of class $\mathcal{R}$. Using results from Chern-Simons theory on hyperbolic three-manifolds and the 3d-3d correspondence we are able to constrain the two independent coefficients in the four-derivative supergravity Lagrangian. This in turn allows us to calculate the subleading terms in the large-$N$ expansion of supersymmetric partition functions for an infinite class of three-dimensional $\mathcal{N}=2$ SCFTs of class $\mathcal{R}$. We also determine the leading correction to the Bekenstein-Hawking entropy of asymptotically AdS$_4$ black holes arising from wrapped M5-branes. In addition, we propose and test some conjectures about the perturbative partition function of Chern-Simons theory with complexified ADE gauge groups on closed hyperbolic three-manifolds.}

\begin{document}

\maketitle

\section{Introduction}
\label{sec:Introduction}

In the absence of a first principle understanding of M-theory calculating higher-derivative (HD) corrections to eleven-dimensional supergravity is a prohibitively hard endeavor. One can resort to employing string/M-theory dualities and access these corrections by computing string scattering amplitudes, see for instance \cite{Green:2005ba}. This is however not an easy task and thus there are few explicit results for the leading higher-derivative corrections in the M-theory low-energy effective action. These HD corrections are of particular interest in the context of holography where they offer calculational access to observables in the dual strongly interacting CFT beyond the leading planar approximation. Recent results in supersymmetric localization offer a new perspective on this old problem and lead to a fruitful interplay with holography and string/M-theory  corrections to supergravity. Taking the gauge/gravity duality as a given, the idea is simple to state. One should find a concrete AdS/CFT dual pair in string/M-theory and then calculate suitable observables in the CFT, for example the partition function on a compact manifold, which can then be mapped to quantities in the gravitational side of the duality. The leading order terms in the large $N$ limit of the CFT observables are then related to the two-derivative terms in the supergravity action. Calculating subleading $1/N$ corrections by supersymmetric localization in the CFT offers a systematic approach to finding the higher-derivative corrections to supergravity. These ideas have been implemented successfully in various contexts over the past few years, see \cite{Chester:2018aca,Binder:2018yvd,Chester:2018dga} in the context of M-theory.

Here, following \cite{Bobev:2020egg}, we explore a related but distinctly different implementation of this program. The idea is to study a four-dimensional HD supergravity theory which encodes the M-theory HD corrections. This can be viewed as an HD extension of the well-known technique of consistent supergravity truncations from 10d/11d to lower-dimensional gauged supergravity. This approach has the technical advantage that the leading HD terms in 4d appear at the four-derivative level which is to be contrasted with the leading eight-derivative correction to 11d supergravity. Indeed, in \cite{Bobev:2020egg,BCHR} it was shown that this idea can be applied successfully for 4d $\mathcal{N}=2$ minimal gauged supergravity where there are two independent four-derivative corrections. Assuming that this 4d $\mathcal{N}=2$ theory is obtained from M-theory compactified on $S^7$ (or an orbifold thereof) and using results for supersymmetric localization of 3d $\mathcal{N}=2$ SCFTs arising on the world-volume of M2-branes it was shown in \cite{Bobev:2020egg,BCHR} how to determine the coefficients of these four-derivative terms. This result can then be used in conjunction with holography to calculate a plethora of observables in the 3d $\mathcal{N}=2$ SCFTs some of which are not accessible by supersymmetric localization. In addition it is possible to calculate explicitly the leading correction to the Bekenstein-Hawking entropy, and other thermodynamic quantities, for any asymptotically AdS$_4$ black hole solution in the theory.

Encouraged by this success it is natural to explore other examples where these ideas can be applied. Our goal in this paper is to show how this can be done for a class of 3d $\mathcal{N}=2$ SCFTs, known as class $\mathcal{R}$, which control the low-energy dynamics of M5-branes wrapped on a hyperbolic 3-manifold $M_3$. It is essential for our discussion that precisely this system of $N$ wrapped M5-branes, in the large $N$ limit, admits a consistent truncation, see \cite{Donos:2010ax}, to the minimal 4d $\mathcal{N}=2$ gauged supergravity  used in \cite{Bobev:2020egg}. The existence of the class $\mathcal{R}$ SCFTs together with this 4d supergravity consistent truncation form the backbone of our analysis. The idea is to combine results from the large $N$ limit of the 3d-3d correspondence initiated in \cite{Dimofte:2011ju} (see \cite{Dimofte:2014ija} for a review and further references) with higher-derivative holographic results to calculate the supersymmetric partition functions on various compact Euclidean manifolds for class $\mathcal{R}$ theories and to gain insight into the microstate counting for supersymmetric AdS$_4$ black holes. This program was implemented successfully in \cite{Gang:2014qla,Gang:2014ema,Gang:2018hjd,Gang:2019uay,Benini:2019dyp,Bobev:2019zmz} in the leading $N^3$ approximations and in some cases for $\log N$ corrections arising from 1-loop contributions in supergravity. Our goal here is to generalize this analysis by using the results in \cite{Bobev:2020egg} and to calculate the subleading order $N$ terms in the large $N$ limit.

More specifically we can use the 3d-3d correspondence to relate the evaluation of the partition function of the 3d $\mathcal{N}=2$ SCFT of class $\mathcal{R}$ on a given 3d supersymmetric background $\mathbb{B}$ to the calculation of the partition function of Chern-Simons theory with a complex gauge group on the hyperbolic manifold $M_3$. In the large $N$ limit the calculation somewhat simplifies and one needs to calculate the Chern-Simons partition function up to two loops. Our main result can be summarized succinctly by the following formula for the partition function of the class $\mathcal{R}$ theory\footnote{The partition function on a Euclidean background is in general a complex quantity.  We focus on the real part of $\log Z_{\mathbb{B}}$ in Sections \ref{sec:HD} and \ref{sec:HoloFBH}, since this is more readily accessible from our Euclidean gravitational theory, although we also make predictions for the imaginary part of the partition function in Section \ref{sec:CS} via the 3d-3d correspondence.\label{foot:real}} 
\begin{equation}\label{eq:logZintro}
-\log Z_{\mathbb{B}} = \frac{{\rm vol}(M_3)}{3\pi}\left[ \mathcal{F}_{\mathbb{B}}\,d_Gh_G + \frac{\chi_{\mathbb{B}}}{4}\,r_G\right]\,.
\end{equation}
Here ${\rm vol}(M_3)$ is the volume of the hyperbolic 3-manifold while $\mathcal{F}$ and $\chi$ are certain geometric quantities associated with a given background $\mathbb{B}$ and summarized in Table~\ref{tab:solutions} for three classes of examples. The dimension, $d_G$, rank, $r_G$, and Coxeter number, $h_G$, of the simply laced Lie algebra $G$ are given in Table~\ref{table:LieA} and the choice of $G$ determines the parent 6d $\mathcal{N}=(2,0)$ SCFT from which the class $\mathcal{R}$ 3d theory descends. The result in \eqref{eq:logZintro} captures the leading and first subleading term in the large $N$ approximation to $\log Z_{\mathbb{B}}$. For $G=A_{N-1}$ the leading $N^3$ term in this expansion  has been discussed before in a holographic context \cite{Gang:2014qla,Gang:2014ema,Gang:2018hjd,Gang:2019uay,Benini:2019dyp,Bobev:2019zmz}. The subleading corrections in \eqref{eq:logZintro} are captured by the four-derivative terms in the 4d $\mathcal{N}=2$ supergravity action. The superconformal index \cite{Bhattacharya:2008zy} and the topologically twisted index \cite{Benini:2015noa} are two choices for $\mathbb{B}$ of particular interest for the physics of asymptotically AdS$_4$ supersymmetric black holes since they can be used to account for the black hole entropy. We show how to apply the result in \eqref{eq:logZintro} in this context and find the first subleading correction to the Bekenstein-Hawking entropy of static magnetic Reissner-Nordstr\"{o}m and rotating Kerr-Newman BPS black holes in AdS$_4$ arising from wrapped M5-branes. As a byproduct of our analysis we find that the validity of \eqref{eq:logZintro} leads to a new result for the scaling with $N$ of the $n$-loop perturbative contribution to the partition function of CS theory on $M_3$ and an explicit prediction for the 2-loop answer. These results may be of interest independently in the study of hyperbolic manifolds and complex CS theory.

In the next section we start with a quick summary of the main results of \cite{Bobev:2020egg,BCHR} on the leading HD corrections to 4d $\mathcal{N}=2$ gauged supergravity. In addition we discuss three classes of explicit solutions of this theory and evaluate their regularized on-shell action. In Section~\ref{sec:HoloFBH} we use these supergravity results together with the 3d-3d results in \cite{Gang:2019uay} to determine the coefficients in the HD supergravity action and derive \eqref{eq:logZintro}. In Section~\ref{sec:CS} we show how the relation in \eqref{eq:logZintro} can be tested using Chern-Simons theory and the 3d-3d correspondence. We conclude in Section~\ref{sec:discussion} with a discussion on some open problems.

\section{Higher-derivative supergravity and its solutions}
\label{sec:HD}

Our supergravity analysis is based on two sets of results. First, we use the fact that 11d supergravity admits a consistent truncation to 4d $\mathcal{N}=2$ minimal supergravity suitable for describing the backreaction of M5-branes wrapping three-manifolds \cite{Donos:2010ax}. The ansatz for the supergravity background fields in this truncation is inspired by an AdS$_4\times H_3$ solution of the maximal 7d $\SO(5)$ gauged supergravity \cite{Pernici:1984xx} found in \cite{Pernici:1984nw} and interpreted as arising from wrapped M5-branes in \cite{Gauntlett:2000ng}. This consistent truncation implies that any solution of the equations of motion of minimal 4d $\mathcal{N}=2$ gauged supergravity can be uplifted to a solution of 11d supergravity. Second, we employ the results in \cite{Bobev:2020egg,BCHR} where it was shown how to construct the four-derivative corrections to 4d $\mathcal{N}=2$ gauged supergravity using conformal supergravity. Combining these two results allows us to study the higher derivative corrections to various supergravity solutions arising from wrapped M5-branes. In order to do this we assume that the consistent truncation results in \cite{Donos:2010ax} can be extended to the four-derivative level. This assumption will be supported by a number of non-trivial consistency checks using the gauge/gravity duality.

As shown in \cite{Bobev:2020egg,BCHR} the four-derivation action of 4d $\mathcal{N}=2$ gauged supergravity has the following bosonic form\footnote{In view of the holographic applications of interest here we work in Euclidean signature. We also note in passing that in this paper we will not consider parity-breaking terms in the supergravity Lagrangian of the form $F\wedge F$ and $R \wedge R$. At the two-derivative level such terms play a role in holography for theories of class $\mathcal{R}$ and will be discussed in \cite{DG-WIP}. Further discussion of these parity-violating terms at the four-derivative level will appear in \cite{BCHR}.\label{foot:parity}}
\begin{equation}
\label{eq:HD-action}
\mathcal{L}_{\text{HD}} = \mathcal{L}_{2\partial} + (c_1 - c_2)\,\mathcal{L}_{\mathrm{W}^2} + c_2\,\mathcal{L}_\mathrm{GB} \, ,
\end{equation}
where
\begin{equation}
\label{eq:HD-action-pieces}
\begin{split}
e^{-1}\mathcal{L}_{2\partial} =&\; -\frac{1}{16\pi\,G_N}\left[R + \frac{6}{L^2} - \frac14\,F_{ab}F^{ab}\right]\, , \\
e^{-1}\mathcal{L}_{\mathrm{W}^2} =&\; \bigl(C_{ab}{}^{cd}\bigr)^2 - \frac{1}{L^2} F_{ab}F^{ab} + \frac12\bigl(F_{ab}^+\bigr)^2\bigl(F_{cd}^-\bigr)^2 \\
&\qquad\qquad\qquad - 4\,F_{ab}^- R^{ac} F^+_c{}^b + 8\,\bigl(\nabla^a F_{ab}^-\bigr)\bigl(\nabla^c F^+_c{}^b\bigr) \, , \\
e^{-1}\mathcal{L}_\text{GB} =&\; R^{abcd}\,R_{abcd} - 4\,R^{ab} R_{ab} + R^2 \, . 
\end{split}
\end{equation}
Here $e^2$ is the determinant of the 4d metric with Riemann and Weyl tensors $R_{abcd}$ and $C_{abcd}$, respectively. $F_{ab}^{\pm}$ are the self-dual and anti-self-dual parts of the graviphoton field strength, $G_N$ is the Newton constant and $L$ determines the AdS$_4$ length scale. The constants $c_1$ and $c_2$ are the coefficients of the only independent four-derivative supersymmetric invariants in $\mathcal{N}=2$,~4d gauged conformal supergravity and together with $L^2/G_N$ they make up the three undetermined dimensionless parameters in the four-derivative bosonic action~\eqref{eq:HD-action}. 

Solving the equations of motion derived from the four-derivative Lagrangian in \eqref{eq:HD-action} is in general a complicated problem. However, it was shown in \cite{Bobev:2020egg,BCHR} that every solution of the two-derivative equations of motion derived from the Lagrangian $\mathcal{L}_{2\partial}$ in \eqref{eq:HD-action-pieces},
\begin{equation}\label{eq:2derEoM}
\begin{split}
R_{\mu\nu} - \frac12\,g_{\mu\nu}\,R - \frac{3}{L^2} g_{\mu\nu} &=  \frac12F_{\mu\rho} F_\nu{}^{\rho}- \frac18\,g_{\mu\nu}\,(F_{\rho\sigma})^2\,, \\
\nabla_{\mu}F^{\mu\nu}&=0\,,
\end{split}
\end{equation}
is also a solution of the full four-derivative equations of motion. In addition, it can be shown that BPS solutions of the two-derivative equations of motion are also supersymmetric in the full four-derivative supergravity theory. These are non-trivial facts which are essential ingredients in the subsequent discussion and we will focus exclusively on such two-derivative solutions.

We are interested in calculating the on-shell action of asymptotically locally AdS$_4$ solutions. It is well-known that this requires adding appropriate boundary terms in order to render the on-shell action finite. To this end we need the following two counterterms to regularize the on-shell action, see \cite{Bobev:2020egg,BCHR} for more details and further references, 
\begin{equation}
\label{eq:CTs}
\begin{split}
I^\text{CT}_{2\partial} =&\; \frac{1}{8\pi\,G_N}\int d^3 x \sqrt{h}\,\left( - K + \tfrac{1}{2}\,L\,\mathcal{R} + \frac{2}{L}\right) \, , \\
I^\text{CT}_{\text{GB}} =&\; 4\int d^3 x \sqrt{h}\,\bigl(\mathcal{J} - 2\,\mathcal{G}_{ab}\,K^{ab}\bigr) \, ,
\end{split}
\end{equation}
where~$h_{ab}$ is the induced metric on the boundary,~$K_{ab}$ is the extrinsic curvature,~$\mathcal{R}$ and $\mathcal{G}_{ab}$ are the boundary Ricci scalar and Einstein tensor, respectively, and~$\mathcal{J}$ is defined by
\begin{equation}
\mathcal{J} = \tfrac13\bigl(3 K (K_{ab})^2 - 2 (K_{ab})^3 - K^3 \bigr) \, .
\end{equation}
Notice that we do not need a separate set of counterterms for the $\mathcal{L}_{\mathrm{W}^2}$ part of the action in \eqref{eq:HD-action-pieces}. This is due to the following \emph{on-shell} relation between the three actions in~\eqref{eq:HD-action-pieces},
\begin{equation}
\label{eq:HD-os-relation}
I_{\mathrm{W}^2} = I_\text{GB} - \frac{64\pi G_N}{L^2}\,I_{2\partial} \, .
\end{equation}
As discussed further in \cite{BCHR} this implies that~\eqref{eq:CTs} provide the complete set of counterterms in order to renormalize the action~\eqref{eq:HD-action} evaluated on a given solution of interest. 

\subsection{Solutions and on-shell action}
\label{sec:Onshell}
  
While a two-derivative solution to~\eqref{eq:2derEoM} is not affected by the four-derivative terms in~\eqref{eq:HD-action}, the corresponding on-shell action is in general modified. It was shown in \cite{Bobev:2020egg,BCHR} that applying the holographic renormalization procedure using the counterterms in \eqref{eq:CTs} leads to the following regularized on-shell action
\begin{equation}\label{eq:Tshirt}
I_{\rm HD} = \left[1+\frac{64\pi G_N}{L^2}(c_2-c_1)\right] \frac{\pi L^2}{2 G_N} \mathcal{F}+32\pi^2 c_1 \chi\,.
\end{equation}
Here $\mathcal{F}$ is the regularized on-shell action of a given solution to the two-derivative supergravity theory\footnote{An efficient way to calculate $\mathcal{F}$ for supersymmetric solutions is to use the results in \cite{BenettiGenolini:2019jdz,Genolini:2016ecx} which express the on-shell action in terms of topological data of the two-derivative solution.} and $\chi$ is the Euler number of the asymptotically AdS$_4$ manifold. We emphasize that this result is valid for any solution of the two-derivative equations of motion including non-supersymmetric solutions. A number of explicit solutions to the equations of motion and the corresponding values of $\mathcal{F}$ and $\chi$ are presented in \cite{Bobev:2020egg,BCHR}. Below we discuss three examples that are of particular interest in the context of wrapped M5-branes and theories of class $\mathcal{R}$.

\subsubsection{AdS-Taub-Bolt}
\label{subsec:AdS-TB}

The first class of solutions we consider is a family of supersymmetric AdS-Taub-Bolt solutions presented in~\cite{Toldo:2017qsh}, see also \cite{AlonsoAlberca:2000cs}. The metric of these solutions reads\footnote{For the explicit solutions of consideration we work in units where the AdS scale is set to~$L = 1$ to simplify the formulas.  The scale is easily restored in the on-shell action via dimensional analysis. \label{foot:TW-units}}
\begin{equation}\label{eq:metBoltpm}
ds^2 = \lambda(r)\,(d\tau + 2s\,f_\kappa(\theta)\,d\phi)^2 + \frac{dr^2}{\lambda(r)} + (r^2 - s^2)\,d\Omega^2_\kappa \, ,
\end{equation}
where
\begin{equation}
\lambda(r) = \frac{(r^2 - s^2)^2 + (\kappa - 4s^2)(r^2 + s^2) - 2\,M\,r + P^2 - Q^2}{r^2 - s^2} \, ,
\end{equation}
and
\begin{equation}
f_\kappa(\theta) = \begin{cases} \cos\theta \quad &\text{for} \quad \kappa = +1 \\ -\theta \quad &\text{for} \quad \kappa = 0 \\ -\cosh\theta \quad &\text{for} \quad \kappa = -1 
\end{cases} \, .
\end{equation}
Here~$\kappa$ is the normalized curvature of a Riemann surface~$\Sigma_\mathfrak{g}$, whose line element in local coordinates reads
\begin{equation}
d\Omega^2_\kappa = \begin{cases} d\theta^2 + \sin^2\theta\,d\phi^2 \quad &\text{for} \quad \kappa = +1 \\ d\theta^2 + d\phi^2 \quad &\text{for} \quad \kappa = 0 \\ d\theta^2 + \sinh^2\theta\,d\phi^2 \quad &\text{for} \quad \kappa = -1 \end{cases} \, .
\end{equation}
In addition to the metric, there is a gauge field with components 
\begin{equation}\label{eq:ABoltpm}
A = \frac{P(r^2+s^2) - 2s\,Q\,r}{r^2 - s^2}\Bigl(\frac{1}{s}\,d\tau + 2\,f_\kappa(\theta)\,d\phi\Bigr) \, .
\end{equation}
The solutions have a mass parameter~$M$, a squashing parameter~$s$, and charge parameters~$(P,Q)$. The radial coordinate is denoted by~$r$ and~$\tau$ parametrizes a circle fibered over the Riemann surface. Asymptotically, the boundary is a smooth 3-manifold~$\mathcal{M}_{\mathfrak{g},p}$ with topology~$\mathcal{O}(-p) \rightarrow \Sigma_\mathfrak{g}$, provided the Euclidean time circle has period
\begin{equation}\label{eq:tauperBolt}
\Delta\tau = \frac{8\pi s}{p}\,|\mathfrak{g} - 1| \quad \text{for}\quad \mathfrak{g} \neq 1 \, , \qquad \Delta\tau = \frac{8\pi s}{p} \quad \text{for} \quad \mathfrak{g} = 1 \, .
\end{equation}
The solutions preserve 1/4 of the supersymmetry of the 4d $\mathcal{N}=2$ supergravity when the parameters obey the relations~\cite{Toldo:2017qsh}
\begin{equation}
P = -\frac12\,(4s^2 - \kappa) \, , \qquad M = 2s\,Q \, .
\end{equation}
These Euclidean solutions are regular in the interior where the circle parametrized by $\tau$ goes smoothly to zero size. Depending on the value of~$Q$ this can happen in one of two ways: either the vanishing locus is a point known as a NUT, or it is a two-dimensional surface called a Bolt. To make contact with the results of~\cite{Gang:2019uay}, we focus on the Bolt-type solutions in what follows. Analyzing the behavior of the function~$\lambda(r)$, it was shown in~\cite{Toldo:2017qsh} that there are actually two distinct classes of Bolt solutions. The first, referred to as Bolt$_+$, is specified by setting
\begin{equation}
Q = Q_+ = \frac{\mathbf{p}^2 - (16s^2 - \mathbf{p})\sqrt{(16s^2 + \mathbf{p})^2 - 128\kappa\,s^2}}{128\,s^2} \, ,
\end{equation}
and the metric is well-defined for~$r \geq r_+ > s$ with
\begin{equation}
r_+ = \frac{\mathbf{p} + \sqrt{(16s^2 + \mathbf{p})^2 - 128\kappa\,s^2}}{16s} \, .
\end{equation}
Here we have introduced the notation
\begin{equation}
\mathbf{p} = \frac{p}{|\mathfrak{g} - 1|} \quad \text{for}\quad \mathfrak{g} \neq 1 \, , \qquad \mathbf{p} = p \quad \text{for} \quad \mathfrak{g} = 1 \, ,
\end{equation}
For the Bolt$_-$ solution on has 
\begin{equation}
Q = Q_- = -\frac{\mathbf{p}^2 - (16s^2 + \mathbf{p})\sqrt{(16s^2 - \mathbf{p})^2 - 128\kappa\,s^2}}{128\,s^2} \, ,
\end{equation}
and the metric is well-defined for~$r \geq r_- > s$ with
\begin{equation}
r_- = \frac{\mathbf{p} - \sqrt{(16s^2 - \mathbf{p})^2 - 128\kappa\,s^2}}{16s} \, .
\end{equation}
The two-derivative on-shell action in~\eqref{eq:HD-action-pieces} receives contributions from the Ricci scalar of the metric, the cosmological constant and the gauge field strengths. The metric and cosmological constant contributions are 
\begin{equation}\label{eq:I2dermetBolt}
I_{2\partial,\,\text{metric}} = \frac{s}{G_N\,\mathbf{p}}\,\Bigl(r_\infty^3 - 3\,s^2\,r_\infty - r_0^3 + 3\,s^2\,r_0\Bigr)\,\text{vol}(\Sigma_\mathfrak{g}) \, ,
\end{equation}
where we have introduced a radial cut-off~$r_\infty$, and~$r_0 = r_+$ or~$r_0 = r_-$ depending on whether we are considering a Bolt$_+$ or Bolt$_-$. The volume factor is given by
\begin{equation}
\text{vol}(\Sigma_\mathfrak{g}) = \begin{cases} 4\pi|\mathfrak{g} - 1| \quad &\text{for} \;\; \mathfrak{g} \neq 1 \\ 4\pi \quad &\text{for} \;\; \mathfrak{g} = 1 \end{cases} \, .
\end{equation}
The expression in \eqref{eq:I2dermetBolt} must be regularized using the two-derivative counterterm in~\eqref{eq:CTs}. This procedure also contributes a finite piece
\begin{equation}
I^\text{finite}_{2\partial,\,\text{CT}} = \frac{2\,s^2}{G_N\,\mathbf{p}}\,Q\,\text{vol}(\Sigma_\mathfrak{g}) \, .
\end{equation}
The graviphoton on-shell action is finite and given by
\begin{equation}
I_{2\partial,\,F} = \frac{r_0\,s}{G_N\,\mathbf{p}}\,\frac{8\,Q\,r_0\,s\,(4s^2 - \kappa) + (r_0^2 + s^2)\,(4\,Q^2 + (4s^2 - \kappa)^2)}{4\,(r_0^2 - s^2)^2}\,\text{vol}(\Sigma_\mathfrak{g}) \, .
\end{equation}
Putting these contributions together and using the values of~$r_0$ and~$Q$ for Bolt$_\pm$ solutions we recover the result of~\cite{Toldo:2017qsh},
\begin{equation}
I_{2\partial} = \frac{\pi}{8\,G_N}\,\bigl(4(1-\mathfrak{g}) \mp p\bigr) \, .
\end{equation}
The Weyl-squared Lagrangian in~\eqref{eq:HD-action-pieces} gives a manifestly finite contribution:\footnote{One also checks that the counterterm needed to renormalize the Weyl-squared action obtained from~\eqref{eq:HD-os-relation} vanishes identically for the Bolt$_\pm$ solutions considered here.} 
\begin{equation}
\label{eq:W2-Bolt_pm}
I_{\mathrm{W}^2} = 8\,\pi^2\,\bigl(4(1 - \mathfrak{g}) \pm p\bigr) \, .
\end{equation}
The Gauss-Bonnet term in~\eqref{eq:HD-action-pieces} has a divergence that needs to be renormalized using the counter term~\eqref{eq:CTs}, which introduces a finite term
\begin{equation}
I^\text{finite}_{\text{GB},\,\text{CT}} = 128\pi\,\frac{s^2}{\mathbf{p}}\,Q\,\text{vol}(\Sigma_\mathfrak{g}) \, ,
\end{equation}
and the total contribution is simply
\begin{equation}
I_{\text{GB}} = 64\,\pi^2\,(1 - \mathfrak{g}) \, .
\end{equation}
Putting all the contributions together with the appropriate coefficients as in~\eqref{eq:HD-action} and restoring the AdS scale $L$, we arrive at the final result for the Euclidean on-shell action, 
\begin{equation}\label{eq:onshellBoltpm}
I_{\text{Bolt}_\pm} = \Bigl[\frac{\pi L^2}{2\,G_N} + 32\,\pi^2\,(c_2 - c_1)\Bigr]\,\frac{4(1 - \mathfrak{g}) \mp p}{4} + 64\,\pi^2\,c_1\,(1 - \mathfrak{g}) \, .
\end{equation}

Some comments are in order. It was emphasized in \cite{Gang:2018hjd} that when this solution is embedded in M-theory and arises from M5-branes wrapping a hyperbolic manifold there is a topological constraint which restricts the allowed values of the integer $p$ to be even, i.e. $p\in 2\mathbb{Z}$. It is clear from \eqref{eq:tauperBolt} that the limit $p=0$ is singular and has to be taken with care. One finds that for $p=s=0$ the solution in \eqref{eq:metBoltpm} and \eqref{eq:ABoltpm} reduces to the so-called Euclidean Romans solutions discussed in detail in \cite{BenettiGenolini:2019jdz,Bobev:2020pjk}, see also \cite{Romans:1991nq,Azzurli:2017kxo}. The on-shell action of this class of Euclidean solutions is given simply by \eqref{eq:onshellBoltpm} for $p=0$. As discussed in \cite{Bobev:2020pjk} one can show that the solutions with $p=s=0$ and $\mathfrak{g}>1$ admit an analytic continuation to a smooth Lorentzian supersymmetric black hole which is simply the extremal AdS-Reissner-Nordstr\"om solution with a hyperbolic horizon.

\subsubsection{Squashed sphere}

Another class of Euclidean solutions that will be of interest here was obtained in~\cite{Martelli:2011fu}. It consists of the following metric and graviphoton (working with~$L=1$ as in Footnote~\ref{foot:TW-units}),
\begin{equation}
\label{eq:U1xU1-sol}
\begin{split}
ds^2 =&\; f_1^2(x,y)\,dx^2 + f_2^2(x,y)\,dy^2 + \frac{(d\Psi + y^2\,d\Phi)^2}{f_1^2(x,y)} + \frac{(d\Psi + x^2\,d\Phi)^2}{f_2^2(x,y)} \, , \\
A =&\; (s^2 - 1)\,\frac{d\Psi - xy\,d\Phi}{(y+x)} \, ,
\end{split}
\end{equation}
where the functions~$f_1, f_2$ are given by
\begin{equation}
f_1^2(x,y) := \frac{y^2 - x^2}{(x^2 - 1)(s^2 - x^2)} \, , \quad f_2^2(x,y) := \frac{y^2 - x^2}{(y^2 - 1)(y^2 - s^2)} \, .
\end{equation}
and the real parameter $s$ obeys $s\geq 1$. The coordinates $x$ and $y$ have the ranges $1\leq x \leq s$ and $s\leq y < \infty$, while the ranges for the angular coordinates $\Psi$ and $\Phi$ are more involved to state and can be found in~\cite{Martelli:2011fu}. These solutions are 1/2-BPS and provide the holographic dual description of a 3d $\mathcal{N}=2$ SCFT on a squashed $S^3$ with a~$U(1)\times U(1)$ isometry \cite{Hama:2011ea}. The parameter~$s$ controls the squashing of the boundary $S^3$ with~$s=1$ corresponding to the round sphere. Indeed, for $s=1$ one finds that the background in \eqref{eq:U1xU1-sol} is simply Euclidean AdS$_4$ in Plebanski-Demianski coordinates. It is straightforward to evaluate the on-shell action for this class of solutions. The two-derivative piece in~\eqref{eq:HD-action-pieces} has a term coming from the metric and the cosmological constant, which gives
\begin{equation}
I_{2\partial,\,\text{metric}} = \frac{3\pi}{2\,G_N}\frac{1}{s(s^2-1)}\int_1^s\,dx\,\int_s^{y_\infty}\,dy\,(y^2 - x^2) \, ,
\end{equation}
where we have introduced a cut-off~$y_\infty$. As shown in~\cite{Martelli:2011fu}, this can be renormalized using the counterterms in~\eqref{eq:CTs}, and the finite piece takes the simple form
\begin{equation}
I^{\text{finite}}_{2\partial,\,\text{metric}} = \frac{\pi}{2\,G_N} \, .
\end{equation}
The two-derivative contribution from the gauge field has a finite action and we can straightforwardly integrate over~$y$ from~$s$ to~$+\infty$. The result is
\begin{equation}
I_{2\partial,\,F} = \frac{\pi}{8\,G_N}\,\frac{(s - 1)^2}{s} \, .
\end{equation}
Putting the two contributions together, we recover the two-derivative regularized on-shell action of~\cite{Martelli:2011fu},
\begin{equation}
I_{2\partial} = \frac{\pi}{2\,G_N}\,\frac{(s+1)^2}{4s} \, .
\end{equation}
Turning to the HD terms, the second and third lines of~\eqref{eq:HD-action-pieces} evaluate on-shell to
\begin{equation}
\label{eq:W2-U1xU1}
e^{-1}\,\mathcal{L}_{\mathrm{W}^2}\big\vert_{\text{o.s.}} = -4\,\frac{(s^2 - 1)^2}{(x+y)^4}\, ,
\end{equation}
and the third line evaluates on-shell to
\begin{equation}
\label{eq:chi-U1xU1}
e^{-1}\,\mathcal{L}_{\text{GB}}\big\vert_{\text{o.s.}} = 24 \, .
\end{equation}
We thus find a contribution to the on-shell action that is manifestly finite from~\eqref{eq:W2-U1xU1}, and one that diverges from~\eqref{eq:chi-U1xU1}. Using~\eqref{eq:CTs} we can renormalize the latter without additional finite contributions.  After then restoring the AdS length scale $L$, we arrive at the result for the renormalized HD Euclidean on-shell action
\begin{equation}
I_{U(1)\times U(1)} = \Bigl[\frac{\pi L^2}{2\,G_N} + 32\,\pi^2\,c_2\Bigr]\,\frac{(s+1)^2}{4s} - 32\,\pi^2\,c_1\,\frac{(s-1)^2}{4s} \, ,
\end{equation}
In the field theory literature it is more common to denote the squashing parameter with~$b$ by the redefinition~$s = b^2$. This leads to the final form of the HD on-shell action,
\begin{equation}\label{eq:U1U1sqI4d}
I_{U(1)\times U(1)} = \Bigl[\frac{\pi L^2}{2\,G_N} + 32\,\pi^2\,(c_2 - c_1)\Bigr]\,\frac14\Bigl(b + \frac{1}{b}\Bigr)^2 + 32\,\pi^2\,c_1 \, ,
\end{equation}
which will be used below in the context of holography.

\subsubsection{AdS-Kerr-Newman}

Finally, we will also consider the 4d AdS-Kerr-Newman (AdS-KN) black hole solution. The metric in Euclidean signature (and with~$L=1$) is given by, see for example~\cite{Caldarelli:1998hg,Cassani:2019mms,Bobev:2019zmz},
\begin{equation}\label{eq:metKN}
ds^2 = \frac{\Delta_r}{W}\Bigl(d\tau + \frac{\alpha}{\Xi}\sin^2\theta d\phi\Bigr)^2 + W\Bigl(\frac{dr^2}{\Delta_r} + \frac{d\theta^2}{\Delta_\theta}\Bigr) + \frac{\Delta_\theta\sin^2\theta}{W}\Bigl(\alpha\,d\tau - \frac{\tilde{r}^2 - \alpha^2}{\Xi}d\phi\Bigr)^2 \, , 
\end{equation}
where
\begin{equation}\label{eq:fnsKN}
\begin{split}
\tilde{r} =&\; r + 2\,m\sinh^2\delta \, , \quad \Xi = 1 + \alpha^2 \, , \quad W(r,\theta) = \tilde{r}^2 - \alpha^2\cos^2\theta \, , \\
\Delta_r(r) =&\; r^2 - \alpha^2 - 2\,m\,r + \tilde{r}^2\,(\tilde{r}^2 - \alpha^2) \, , \quad \Delta_\theta(\theta) = 1 + \alpha^2\cos^2\theta \, .
\end{split}
\end{equation}
The gauge field is given by
\begin{equation}\label{eq:AKN}
A = 2i\,m\sinh(2\delta)\,\frac{\tilde{r}}{W}\Bigl(d\tau + \frac{\alpha}{\Xi}\sin^2\theta\,d\phi\Bigr) \, ,
\end{equation}
where the factor of~$i$ is due to the fact that we work in Euclidean signature. This solution preserves 1/4 of the supersymmetry when $\alpha$ obeys the constraint
\begin{equation}\label{eq:susyrelKN}
\alpha = \frac{2i}{e^{4\delta} - 1} \, .
\end{equation}
As shown in~\cite{Cassani:2019mms,Bobev:2019zmz}, the two-derivative regularized on-shell action can be written as
\begin{equation}
I_{2\partial} = \frac{\pi}{2\,G_N}\,\frac{(\omega + 1)^2}{2\,\omega} \, , 
\end{equation}
where, denoting~$c = \cosh\delta$ and~$s = \sinh\delta$ for brevity,
\begin{equation}\label{eq:omUpsdef}
\omega = -\frac{4}{\Upsilon}\,c\,s\,\bigl(c(c - 2\,s) + s(s - 2\,c)\bigr)\, , \quad \Upsilon = -16\,i\,R_+\,c^2\,s^2 - \sinh(4\delta) \, .
\end{equation}
The parameter~$R_+$ is related to the location of the outer horizon\footnote{In Euclidean signature $r_+$ is simply the value of the radial coordinated at which the space smoothly caps off.}~$r_+$ as~$R_+ = r_+ + 2\,m\,s^2$, and thus the parameter~$m$ can be expressed in terms of~$R_+$ as
\begin{equation}
m = i\,\frac{R_+{}^2 + 1 - (1 + i\,R_+)\coth(2\delta)}{2\,s\,c} \, .
\end{equation}
The parameter~$\delta$ controls the electric charge of the AdS-KN solutions. The AdS-Kerr solution is obtained by setting~$\delta = 0$. 

With this at hand we can compute the regularized on-shell Gauss-Bonnet action using the counterterms in~\eqref{eq:CTs}. A lengthy computation yields the simple result
\begin{equation}
I_{\text{GB}} = 64\,\pi^2 \, .
\end{equation}
Finally, using~\eqref{eq:HD-os-relation} and additionally restoring the AdS length scale $L$, we obtain the following regularized HD Euclidean on-shell action of the supersymmetric AdS-KN solution,
\begin{equation}\label{eq:AdSKNI4d}
I_{\text{KN}} = \Bigl[\frac{\pi L^2}{2\,G_N} + 32\,\pi^2\,(c_2 - c_1)\Bigr]\,\frac{(\omega + 1)^2}{2\,\omega} + 64\,\pi^2\,c_1 \, .
\end{equation}
We note that there is a two-parameter family of supersymmetric Euclidean solutions labelled by $(\delta,m)$ but the on-shell action depends only the specific combination of these parameters given by $\omega$  in \eqref{eq:omUpsdef}.

The Euclidean KN supersymmetric solution presented above can be analytically continued into a regular supersymmetric Lorentzian black hole solution by setting
\begin{equation}
\alpha = i \mathfrak{a} \, ,
\end{equation}
which amounts to a Wick-rotation. This Lorentzian solution is smooth and free of CTCs only if one further relates the mass and rotation parameters as 
\begin{equation}\label{eq:maKNBH}
m=\mathfrak{a}(1+\mathfrak{a})\sqrt{2+\mathfrak{a}}\,,
\end{equation}
and restricts the rotation parameter to lie in the range $0\leq \mathfrak{a} < 1$. We further note that the energy, charge and angular momentum of the black hole take the form
\begin{equation}\label{eq:EQJKN}
E = \frac{m}{G_N \Xi^2} \cosh(2\delta)\,, \qquad Q = \frac{m}{G_N \Xi^2} \sinh(2\delta)\,, \qquad J = \frac{m\mathfrak{a}}{G_N \Xi^2} \cosh(2\delta)\,, 
\end{equation}
and in the BPS limit these quantities obey the relations
\begin{equation}
E = J+Q\,, \qquad J = \frac{Q}{2}\left[\sqrt{1+4G_N^2Q^2}-1\right]\,.
\end{equation}
%

\section{Holographic free energy and black hole entropy}
\label{sec:HoloFBH}

We now proceed to embed the results above in a concrete M-theory setup given by M5-branes wrapping a hyperbolic 3-manifold $M_3$ and study the relation with the low-energy 3d $\mathcal{N}=2$ dual SCFT. For concreteness we first focus on $N$ M5-branes with a transverse flat space for which the low-energy six-dimensional physics is described by the $\mathcal{N}=(2,0)$ SCFT of type $A_{N-1}$.

To relate the parameters in the four-dimensional supergravity theory to quantities in the dual SCFT we follow the same logic as in  \cite{Bobev:2020egg,BCHR}. Based on the AdS/CFT dictionary, we expect the following relation\footnote{An order $N^2$ term in this relation could in principle be generated but is absent when the 3-manifold $M_3$ is compact and smooth. This is analogous to the holographic description of 4d $\mathcal{N}=2$ SCFTs of class $\mathcal{S}$ where an $N^2$ term in the conformal anomaly coefficients is absent when there are no punctures on the Riemann surface \cite{Gaiotto:2009gz}.} 
\begin{equation}\label{eq:L2GN}
\frac{\pi L^2}{2 G_N} = \frac{\text{vol}(M_3)}{3\pi}\,\,N^3 + \pi\,a\,N \, ,
\end{equation}
where~$a$ is an unknown constant that does not depend on~$N$. The coefficient of the $N^3$ term in \eqref{eq:L2GN} has been computed in \cite{Gang:2014qla} and can be deduced from the 7d maximal supergravity or M-theory embedding of the AdS$_4$ vacuum of the 4d $\mathcal{N}=2$ gauged supergravity. The supergravity couplings $c_{1,2}$ in \eqref{eq:HD-action} are expected to scale as $N$ and for convenience we introduce the reparametrization
\begin{equation}
c_i = \frac{v_i}{32\pi} N\,, \qquad i=1,2\,,
\end{equation}
where $v_{i}$ do not scale with $N$. We note in passing that by studying the spectrum of excitations around the AdS$_4$ vacuum solution one can show, see \cite{BCHR}, that in order to avoid superluminal propagation, or alternatively to ensure unitarity of the dual CFT, the inequality $c_2<c_1$ has to be obeyed.

\begin{table}[ht]
	\begin{center}
	\setlength{\tabcolsep}{7pt}
	\def\arraystretch{1.3}
	\begin{tabular}{ c | c | c | c } \hline
	{Solution} & Supersymmetry	 & ${\cal F}$ & $\chi$ \\ \hline \hline
	{Bolt$_{\pm}$} & 1/4 &  $(1-\mathfrak{g}) \mp \frac{\mathfrak{}p}{4}$ & $2(1-\mathfrak{g})$ \\ \hline
	{$U(1) \times U(1)$ sq.} & 1/2 &  $ \frac14 ( b + \frac{1}{b})^2 $ & $1$ \\ \hline
	{KN-AdS} & 1/4 &  $ \frac{(\omega + 1)^2}{2\omega} $ & $2$ \\ \hline
	\end{tabular}
	\end{center}
	\caption{The values of $(\mathcal{F},\chi)$ appearing in the on-shell action~\eqref{eq:Tshirt} for the three classes of Euclidean solutions presented in Section~\ref{sec:Onshell}. }
	\label{tab:solutions}
\end{table}

With this at hand and using \eqref{eq:Tshirt} one finds the following form for the leading and subleading terms in the large $N$ expansion of the partition function of the 3d $\mathcal{N}=2$ SCFT on a manifold with given $\mathcal{F}$ and $\chi$.
\begin{equation}\label{eq:tshirtQFT}
-\log Z = \pi\mathcal{F}\left(A N^3+B N\right) - \pi (\mathcal{F}-\chi) C N\,.
\end{equation}
Here we have defined yet another set of constants
\begin{equation}\label{eq:ABCdef}
A = \frac{\text{vol}(M_3)}{3\pi^2}\,, \qquad B = a + v_2\,, \qquad  C = v_1\,.
\end{equation}
We can fix the constants $(B,C)$ for the theories of class $\mathcal{R}$ on~$M_3$ using the first line in Equation (4.13) of the published version of \cite{Gang:2019uay}. It reads
\begin{equation}
\label{eq:logZATB}
\log Z_{\mathcal{M}_{\mathfrak{g},p}} = \frac{4(\mathfrak{g}-1)+p}{12\pi} {\rm vol}(M_3) (N^3-N) + \frac{(\mathfrak{g}-1)}{6\pi} {\rm vol}(M_3) (N-1)\,. 
\end{equation}
This result should be compared to the on-shell action of the Bolt$_+$ solution, see \eqref{eq:onshellBoltpm} and Table~\ref{tab:solutions}, and we should use the standard AdS/CFT dictionary
\begin{equation}
I_{\text{Bolt}_+} = - \log Z_{\mathcal{M}_{\mathfrak{g},p}}\,.
\end{equation}
Using this result we can fix the constants $B$ and $C$ controlling the order $N$ terms in \eqref{eq:tshirtQFT}
\begin{equation}\label{eq:ABCclassR}
B= - \frac{1}{4\pi^2} {\rm vol}(M_3) \,, \qquad C= \frac{1}{12\pi^2} {\rm vol}(M_3)\,.
\end{equation}
This in turn allows us to find the explicit form of the leading and subleading terms in $\log Z$ for any three manifold which admits a smooth 4d supergravity dual solution by using \eqref{eq:tshirtQFT}. In Table II of \cite{Bobev:2020egg} the values of $(\mathcal{F},\chi)$ for a number of known supergravity solutions are tabulated. Of particular interest to the discussion here are the free energy for the squashed sphere and the superconformal index which we now discuss in some detail. For the squashed sphere we can use \eqref{eq:U1U1sqI4d} and the results above to find a holographic answer for the squashed sphere free energy of the 3d $\mathcal{N}=2$ class $\mathcal{R}$ SCFT
\begin{equation}\label{eq:logZsqS}
F_{S^3_b}=-\log Z_{S^3_b} = \frac{1}{4}\left(b+\frac{1}{b}\right)^2 \frac{{\rm vol}(M_3)}{3\pi} (N^3-N) + \frac{{\rm vol}(M_3)}{12\pi} N\,. 
\end{equation}
The superconformal index is captured holographically by the supersymmetric Euclidean Kerr-Newman solution. Therefore, we can use the on-shell action in \eqref{eq:AdSKNI4d} and the results above to find the following holographic prediction for the leading and subleading terms in the index
\begin{equation}\label{eq:FKN}
F_{S^1\times_{\omega}S^2}=-\log Z _{S^1\times_{\omega}S^2}= \frac{(\omega+1)^2}{2\omega} \frac{{\rm vol}(M_3)}{3\pi} (N^3-N) + \frac{{\rm vol}(M_3)}{6\pi} N\,. 
\end{equation}
As we discuss in Section~\ref{sec:CS} below the results in \eqref{eq:logZsqS} and \eqref{eq:FKN} can be confirmed explicitly on the QFT side by using the 3d-3d correspondence.

Another quantity of interest that can be computed from the results above is the coefficient $C_T$ in the stress-tensor two-point function. It can be obtained by taking two-derivatives of the squashed $S^3$ free energy with respect to the squashing parameter $b$
\begin{equation}\label{eq:CTderFSb}
C_T = \frac{32}{\pi^2} \frac{\partial^2 F_{S^3_b}}{\partial b^2}|_{b=1}\,.
\end{equation}
Using \eqref{eq:logZsqS} we thus find 
\begin{equation}
C_T = \frac{64}{3\pi^3}\frac{{\rm vol}(M_3)}{3\pi} (N^3-N)\,.
\end{equation}
We emphasize that this is a new supergravity prediction for the two-point function of the stress-tensor in 3d $\mathcal{N}=2$ class $\mathcal{R}$ SCFTs. We currently do not know of a CFT method to compute this quantity.

Given the results for the class $\mathcal{R}$ 3d SCFTs arising from the $A_{N-1}$ $\mathcal{N}=(2,0)$ 6d theory and the structure of the 6d $\mathcal{N}=(2,0)$ anomaly polynomial, see \cite{Witten:1996hc,Harvey:1998bx,Yi:2001bz,Intriligator:2000eq}, we can conjecture the following results for the free energy\footnote{These results are for the real part of the free energy, as explained in Footnotes~\ref{foot:real} and~\ref{foot:parity}.} of a general class $\mathcal{R}$ 3d SCFTs arising from an ADE 6d $\mathcal{N}=(2,0)$ SCFT
\begin{equation}\label{eq:largeN}
\begin{split}
F_{\mathcal{M}_{\mathfrak{g},p}} &= -\log Z = \frac{4(1-\mathfrak{g})-p}{4} \frac{{\rm vol}(M_3)}{3\pi}\,d_Gh_G + \frac{(1-\mathfrak{g})}{2} \frac{{\rm vol}(M_3)}{3\pi}\,r_G\,,\\
F_{S^3_b}&=-\log Z = \frac{1}{4}\left(b+\frac{1}{b}\right)^2 \frac{{\rm vol}(M_3)}{3\pi}\,d_Gh_G + \frac{1}{4} \frac{{\rm vol}(M_3)}{3\pi}\,r_G\,,\\
F_{S^1\times_{\omega}S^2}&=-\log Z = \frac{(\omega+1)^2}{2\omega} \frac{{\rm vol}(M_3)}{3\pi}\,d_Gh_G + \frac{1}{2}\frac{{\rm vol}(M_3)}{3\pi}\,r_G\,.
\end{split}
\end{equation}
Here $(d_G,h_G,r_G)$ are the dimension, Coxeter number, and rank of a simply-laced Lie algebra, see Table~\ref{table:LieA}. Using \eqref{eq:CTderFSb} we can also find a conjectural expression for $C_T$
\begin{equation}
C_T = \frac{64}{3\pi^3}\frac{{\rm vol}(M_3)}{3\pi}\,d_G h_G\,.
\end{equation}
For the $D_N$ series the expressions above should be treated as the leading order terms in a large $N$ expansion. For the $E_{6,7,8}$ exceptional series the meaning of these equations is less clear since~$(d_G,h_G,r_G)$ do not scale with~$N$. In Section~\ref{sec:CS} some non-trivial evidence for the validity of \eqref{eq:largeN} will be presented by leveraging results from the 3d-3d correspondence. Finally, we note that the result in \eqref{eq:tshirtQFT} can be used also for other 3d supersymmetric partition functions, such as the ones discussed in \cite{Closset:2017zgf,Closset:2019hyt}, that can be computed by supersymmetric localization. 

\begin{table}[h]
\center{
    \begin{tabular}{|c|c|c|c|}
    \hline
    $G$ & $r_G$ & $d_G$ & $h_G$\\
    \hline
    $A_{N-1}$ & $N-1$ & $N^2-1$ & $N$ \\
    $D_{N}$ & $N$ & $2N^2-N$ & $2N-2$\\
    $E_6$ & $6$ & $78$ & $12$\\
    $E_7$ & $7$ & $133$ & $18$\\
    $E_8$ & $8$ & $248$ & $30$\\
     \hline
    \end{tabular}
    }
      \caption{Simply laced Lie Algebras and their rank $r_G$, dimension $d_G$, and Coxeter number $h_G$.}\label{table:LieA}
\end{table}

We now change gears with a brief discussion on black hole entropy in this class of models. As shown in \cite{Bobev:2020egg,BCHR} the entropy for two-derivative asymptotically AdS$_4$ black hole solutions of the action in \eqref{eq:HD-action} is given by
\begin{equation}\label{eq:SBHtshirt}
S = \left(1+\frac{64 \pi G_N}{L^2}(c_2-c_1)\right) \frac{\mathcal{A}_H}{4G_N} -32\pi^2 c_1 \chi(H)\,.
\end{equation}
Here $\mathcal{A}_H$ is the area of the horizon and $\chi(H)$ is its Euler characteristic. 

We have mainly focused on  Euclidean solutions in Section~\ref{sec:Onshell}. However two of the backgrounds presented there admit analytic continuation to Lorentzian black holes with regular horizons. We can therefore embed these black holes in M-theory and treat them as arising from M5-branes wrapped on the 3-manifold $M_3$ in order to gain insight into their microscopic structure.

As discussed below \eqref{eq:onshellBoltpm}, for $p=s=0$ and $\mathfrak{g}>1$ the solution in \eqref{eq:metBoltpm} and \eqref{eq:ABoltpm} becomes the extremal AdS-Reissner-Nordstr\"om solution with a hyperbolic horizon. The two-derivative Bekenstein-Hawking entropy of this solution reads
\begin{equation}
S_{\rm Romans}^{2\partial} = (\mathfrak{g}-1) \frac{\pi L^2}{2G_N} = (\mathfrak{g}-1) \frac{{\rm vol}(M_3)}{3\pi} N^3\,.
\end{equation}
With this at hand we can use \eqref{eq:SBHtshirt} and the results for $c_{1,2}$ above to find that the first subleading correction to this entropy reads 
\begin{equation}\label{eq:SRomasnHD}
S_{\rm Romans} = [A N^3+(B-C)N]\pi (\mathfrak{g}-1) +2 \pi C N (\mathfrak{g}-1)\,.
\end{equation}
Here we need to use the constants $(A,B,C)$ as in \eqref{eq:ABCdef} and \eqref{eq:ABCclassR}.

The supersymmetric limit of the Kerr-Newman black hole is discussed around \eqref{eq:maKNBH}. Its two-derivative Bekenstein-Hawking entropy can be readily calculated and is given by, see for instance \cite{Bobev:2019zmz}, 
\begin{equation}
S_{\rm KN-AdS}^{2\partial} =  \left[ \sqrt{1+4G_N^2 Q^2}-1\right] \frac{\pi L^2}{2G_N} = \left[ \sqrt{1+4G_N^2 Q^2}-1\right]\frac{{\rm vol}(M_3)}{3\pi} N^3\,,
\end{equation}
where we have used the expression for the charge of the black hole, $Q$, in \eqref{eq:EQJKN}. We can then again employ \eqref{eq:SBHtshirt}, in conjunction with the results above, to find the following leading order correction to this entropy
\begin{equation}\label{eq:SKNHD}
S_{\rm KN-AdS} = [A N^3+(B-C)N]\pi \left[ \sqrt{1+4G_N^2 Q^2}-1\right] -2 \pi C N\,,
\end{equation}
where again we should use \eqref{eq:ABCdef} and \eqref{eq:ABCclassR}. We note that when applying \eqref{eq:SBHtshirt} we have used that for these two classes of black holes the Euler characteristic of the horizon, $\chi(H)$, is actually equal to the Euler characteristic, $\chi$, of the full four-dimensional Euclidean solution given in Table~\ref{tab:solutions}, see \cite{Bobev:2020egg,BCHR}. We conclude by emphasizing that while we have presented explicit expressions for the leading correction to the Bekenstein-Hawking entropy for two specific black hole solutions the result in \eqref{eq:SBHtshirt} can be applied to any asymptotically AdS$_4$ black hole of interest, including non-supersymmetric ones, that solves the equations of motion of the 4d $\mathcal{N}=2$ gauged supergravity.

To finish this discussion we would like to emphasize an important conceptual point. The expressions for the black hole entropy in \eqref{eq:SRomasnHD} and \eqref{eq:SKNHD} can be obtained from the free energy in \eqref{eq:logZATB} (for $p=0$) and \eqref{eq:FKN}, respectively. This can be done by using the standard thermodynamic relations between entropy and free energy. Notice however, that the free energies in \eqref{eq:logZATB} and \eqref{eq:FKN} are computed as a regularized on-shell action for a more general class of Euclidean solutions not all of which have a Lorentzian continuation as regular black holes. More specifically, note that \eqref{eq:FKN} is valid for the 2-parameter family of Euclidean solutions in \eqref{eq:metKN}, \eqref{eq:fnsKN}, \eqref{eq:AKN} with the supersymmetry constraint in \eqref{eq:susyrelKN}. The regular BPS Lorentzian black hole solution, with the entropy given in \eqref{eq:SKNHD}, however, exists only for a one-parameter subfamily of these solutions specified by the relation in \eqref{eq:maKNBH}. The dependence of the free energy in \eqref{eq:FKN} only on the single parameter $\omega$ for the whole 2-parameter family of solutions is a reflection of the fact that in the dual SCFT the other independent combination of parameters is Q-exact with respect to the localization supercharge and therefore does not affect the supersymmetric partition function. For the $p=0$ solutions in Section~\ref{subsec:AdS-TB} this point was also discussed in greater detail in \cite{Bobev:2020pjk}.

\section{Results from the 3d-3d correspondence and Chern-Simons theory}
\label{sec:CS}

The supergravity and holographic results in the previous sections can be understood from an alternative vantage point by using their dual QFT description. To this end we take the 6d $\mathcal{N}=(2,0)$ theory of type $G$ and place it on $\mathbb{R}^3\times M_3$ where $M_3$ is a compact hyperbolic 3-manifold. To preserve 3d $\mathcal{N}=2$ supersymmetry on $\mathbb{R}^3$ we perform a partial topological twist by turning on a background gauge field for an $\SO(3)$ subgroup of the $\SO(5)$ R-symmetry of the 6d theory. In the IR, i.e. at length-scales much larger than the size of $M_3$, we have a 3d $\mathcal{N}=2$ QFT with a $\rm U(1)$ R-symmetry which we refer to as $T[M_3, G]$. This procedure can be summarized schematically as follows 
\begin{align}
6d\; \mathcal{N}=(2,0)\;\textrm{theory of type } G  \textrm{ on } \mathbb{R}^{3}\times M_3 \xrightarrow{\quad \textrm{size}(M_3)\rightarrow 0 \quad } 3d \;T[M_3, G] \textrm{ theory on }\mathbb{R}^{3}\;. \notag
\end{align}
The study of these 3d theories of class $\mathcal{R}$ was initiated in \cite{Dimofte:2011ju}, see \cite{Dimofte:2014ija} for a review, where it was shown that some of their physical observables can be related to observables in Chern-Simons theory with a complexified gauge group $G$ on the manifold $M_3$. This so-called 3d-3d correspondence plays a central role in our discussion below. In particular we employ the fact that we can place the $T[M_3, G]$ theory in a 3d supersymmetric background $\mathbb{B}$. The supersymmetric partition function on this background is then related to a particular topological invariant of the $G_{\mathbb{C}}$ Chern-Simons theory on $M_3$. Schematically this 3d-3d relation reads
\begin{align}
\textrm{3d-3d relation : } Z\left[T[M_3, G] \textrm{ on }\mathbb{B}\right] = ({\rm Invariant}_{\mathbb{B}} \textrm{ of }G_{\mathbb{C}} \textrm{ Chern-Simons theory on }M_3)\;. \notag
\end{align}
Below, we present some explicit expressions for the ${\rm Invariant}_{\mathbb{B}}$ for various supersymmetric backgrounds $\mathbb{B}$. Since we are ultimately interested in taking the large $N$ limit of the rank of the gauge group we first discuss some perturbative results for $G_{\mathbb{C}}$ Chern-Simons theory.

\subsection{Perturbative Chern-Simons theory} 

The perturbative invariants of  Chern-Simons theories with gauge group $G_{\mathbb{C}}$ are determined by flat $G_{\mathbb{C}}$-connections $\mathcal{A}^\alpha_G$ and their corresponding $n$-loop effective actions $S_n^\alpha[M_3, G_{\mathbb{C}}]$.  A flat connection $\mathcal{A}^\alpha$ obeys $d\mathcal{A}^\alpha + \mathcal{A}^\alpha \wedge \mathcal{A}^\alpha =0$ and is fully characterized by its holonomy matrices, $\rho^\alpha (a) := P \exp \left( \oint_a \mathcal{A}^\alpha \right)$ with $a\in \pi_1 (M_3)$. Therefore a $G_{\mathbb{C}}$-flat connection $\mathcal{A}^\alpha$ can be regarded as a homomorphism $\rho^\alpha$ from $\pi_1 (M_3)$ to $G_{\mathbb{C}}$. To this end let us define
\begin{align}
\begin{split}
\chi_{\rm irred} (M_3, G_{\mathbb{C}}) &:= (\textrm{the set of irreducible flat $G_{\mathbb{C}}$-connections on $M_3$})
\\
&=\frac{\textrm{Hom}_{\rm irred}\left[\pi_1 M_3 \rightarrow G_{\mathbb{C}}\right]}{\textrm{(conjugation)}}\;.
\end{split}
\end{align}
Here $\textrm{Hom}_{\rm irred}\left[\pi_1 (M_3) \rightarrow G_{\mathbb{C}}\right]$ is the set of irreducible homomorphism from $\pi_1 (M_3)$ to $G_{\mathbb{C}}$. A homomorphism $\rho : \pi_1 (M_3)\rightarrow G_{\mathbb{C}}$ is called irreducible if 
\begin{align}
\textrm{dim}\; \textrm{\bf Stab}(\rho) =0\;\; \textrm{where }  \textrm{\bf Stab}(\rho)  := \{h\in G_{\mathbb{C}}\;: \; [h,\rho(a)]=0\; \forall a \in \pi_1 (M_3) \} \;.
\end{align}
The perturbative invariants $S^\alpha_n $ associated to a flat connection $\mathcal{A}^\alpha$ are defined as a formal perturbative expansion of the path-integral 
\begin{equation}\label{eq:CSpathint}
\int \frac{D(\delta \mathcal{A})}{(\textrm{gauge})} e^{- \frac{4\pi^2}{\hbar} CS[\mathcal{A}^\alpha +\delta \mathcal{A};M_3, G_{\mathbb{C}}]}\xrightarrow{\quad \hbar \rightarrow 0 \quad }  \frac{1}{\textrm{vol}(\textrm{\bf Stab} (\mathcal{A}^\alpha))} \exp \left(\sum_{n=0}^{\infty} \hbar^{n-1}S_n^\alpha \right) \;,
\end{equation}
where the classical Chern-Simons action is
\begin{equation}
CS[\mathcal{A};M_3,G_{\mathbb{C}}]:= \frac{1}{8\pi^2} \int_{M_3} \textrm{Tr}_G \left( \mathcal{A}\wedge d\mathcal{A}+ \frac{2}3 \mathcal{A}^3 \right)\;.
\end{equation}
Note that the perturbative expansion vanishes for reducible homomorphism $\rho$ due to the volume factor $1/\textrm{vol}({\mathbf{Stab}}(\rho))=1/\infty$. We use the following normalization for the non-degenerate symmetric bilinear form $\textrm{Tr}_G$ on the Lie algebra $ \textrm{Lie}(G)$ of $G$
\begin{align}
{\rm Tr}_{G}(\beta^2)=2 \;\textrm{  for all long roots }  \beta \in \textrm{Lie}(G)\;. 
\end{align}
The leading term in the perturbative expansion in \eqref{eq:CSpathint} is $S_0^\alpha$ which is related to the classical CS action
\begin{align}
S_0^\alpha [M_3, G_{\mathbb{C}}]= -4\pi^2 CS[\mathcal{A}^\alpha;M_3, G_{\mathbb{C}}]\;.
\end{align}
The first subleading term $S_1^\alpha$ is the 1-loop contribution to the path integral and is related to a well-known object in the mathematics of 3-manifolds
\begin{align}\label{1-loop invariant}
S_1^\alpha [M_3, G_{\mathbb{C}}] = - \frac{1}2 \log  \left( \textrm{Tor}_{\rm Adj}[\mathcal{A}^\alpha;M_3, G_{\mathbb{C}}] \right)\;.
\end{align}
Here 
$\textrm{Tor}_{\rm Adj}[\mathcal{A}^\alpha;M_3,G_{\mathbb{C}}]$ is the  Ray-Singer-Reidemeister torsion in the $R=(\rm adjoint)$ representation twisted by the flat $G_{\mathbb{C}}$-connection $\mathcal{A}^\alpha$ \cite{reidemeister1935homotopieringe,ray1971r,cheeger1977analytic,muller1978analytic}. The invariant in the general representation $R\in \textrm{Hom}[G_\mathbb{C}\rightarrow GL(V_R)]$ can be defined as follows
\begin{align}\label{Def : Torsion}
\textrm{Tor}_{R} [\mathcal{A};M_3, G_{\mathbb{C}} ] :=  \frac{[\det' \Delta_0 (R, \mathcal{A})]^{3/2}}{[\det' \Delta_1 (R, \mathcal{A})]^{1/2}}\;.
\end{align}
Here $\Delta_n (R, \mathcal{A})$  is a Laplacian acting on $V_R$-valued $n$-form twisted by a flat connection $\mathcal{A}$ and  $\det' $ is the zeta function regularized determinant. In the definition, we need to choose a metric structure on $M_3$ to define Laplacian operators but the final $\textrm{Tor}_{R}$ is independent on the choice. When $R= ({\rm adjoint})$, the torsion is related to the 1-loop invariant $S_1^\alpha$ of $G_\mathbb{C}$ Chern-Simons theory as given in \eqref{1-loop invariant}. 

For hyperbolic 3-manifolds, there are two special irreducible flat $G_{\mathbb{C}}$-connections $\mathcal{A}^{\rm geom}_{G}$ and its complex conjugate $\mathcal{A}^{\overline{\rm geom}}_G$ which can be constructed from the unique hyperbolic structure on $M_3$~\cite{Garoufalidis:2015ula}.\footnote{For simplicity, we assume that the cohomology $H^1(M, Z_G)$ is trivial where $Z_G$ is the center group of $G$. In general there are $|H^1(M, Z_G)|$ many copies of $\mathcal{A}^{\rm geom}_{G}$ and $\mathcal{A}^{\overline{\rm geom}}_G$ related to each other by the tensoring with $Z_G$ flat connections.} These flat connections can be constructed using the principal embedding $\rho^G_{\rm pr} : SU(2)\rightarrow G$,
\begin{align}\label{eq:Ageomdef}
\mathcal{A}_G^{\rm geom} = \rho^G_{\rm pr} \cdot (\omega+ i e)\;, \quad  \mathcal{A}_G^{\overline{\rm geom}} = \rho^G_{\rm pr} \cdot (\omega- i e)\;.
\end{align}
Here $\omega$ and $e$ are the spin-connection and dreibein for the unique hyperbolic metric on $M_3$, normalized as $R_{\mu\nu} = -2g_{\mu\nu}$, and can be regarded as $SO(3)$-valued 1-forms on $M_3$. Their complex combination $\omega + i e $ defines a flat $SO(3,\mathbb{C}) = PSL(2,\mathbb{C}) = SL(2,\mathbb{C})/\mathbb{Z}_2$-connection which can always be uplifted to an $SL(2,\mathbb{C})$ flat connection~\cite{Witten:1988hc,Gukov:2003na}. The principal embedding $\rho_{\rm pr}^G$ is defined by the following branching rules for the adjoint representation of the simply laced algebra $G$ 
 \begin{align}
 \begin{split}
 &\mathbf{N^2-1}~~{\rm of } \;A_{N-1}  \rightarrow  \bigoplus_{m=1}^{N-1}\tau_{2 m+1}\;,
 \\
  &\mathbf{2N^2-N}~~{\rm  of } \;D_{N} \rightarrow  \left(\bigoplus_{m=1}^{N-1}\tau_{4 m-1} \right) \oplus \tau_{2N-1}\;,
  \\
 &\mathbf{78}~~{\rm  of } \;E_{6}   \rightarrow \tau_3\oplus\tau_9\oplus \tau_{11} \oplus \tau_{15} \oplus \tau_{17} \oplus \tau_{23}   \;, 
    \\
 &\mathbf{133}~~{\rm of } \;E_{7}  \rightarrow  \tau_3 \oplus \tau_{11}  \oplus \tau_{15} \oplus \tau_{19} \oplus \tau_{23} \oplus \tau_{27} \oplus \tau_{35}    \;,
      \\  
  &\mathbf{248}~~{\rm of } \;E_{8} \rightarrow  \tau_3 \oplus \tau_{15} \oplus \tau_{23} \oplus \tau_{27} \oplus \tau_{35} \oplus \tau_{39} \oplus \tau_{47} \oplus \tau_{59}   \;. 
 \end{split} \label{branching rule}
 \end{align}
Here $\tau_{n}$ is the $n$-dimensional unitary irreducible representation of $SU(2)$.  These flat connection enjoy the following inequalities \cite{Dimofte:2013iv}
\begin{align}
\textrm{Im} \left( CS[\mathcal{A}^{\overline{\rm geom}}_G ]\right)\; > \;\textrm{Im} \left( CS[\mathcal{A}^{\alpha } ]\right) \; >\; \textrm{Im} \left( CS[\mathcal{A}^{\rm geom}_G ]\right)\;, \label{property-of-A-geom}
\end{align}
valid for all other irreducible $G_{\mathbb{C}}$ flat connections $\mathcal{A}^{\alpha }$. Using \eqref{eq:Ageomdef} one can show that the classical perturbative invariants $S^\alpha_0$ for $\alpha = (\textrm{geom})$ and  $(\overline{\textrm{geom}}) $ are given by
\begin{align}
\begin{split}
& S_0^{\rm geom}[M_3, G_\mathbb{C}] = \left( S_0^{\overline{\rm geom}}[M_3, G_\mathbb{C}]\right)^*  = \textrm{Ind}(\rho_{\rm pr}^G ) \times S_0^{\rm geom}[M_3, SU(2)_\mathbb{C}]\;.
\end{split}
\end{align}
Here  the group theoretical factor $\textrm{Ind}(\rho_{\rm pr}^G ) $ is defined as
\begin{align}
\textrm{Tr}_G \left(\rho^G_{\rm pr}(h_1) \rho^G_{\rm pr} (h_2)\right) = \textrm{Ind}(\rho_{\rm pr}^G ) \times  \textrm{Tr}_{SU(2)} (h_1 h_2)\;, \quad \textrm{for all } h_1, h_2 \in \mathfrak{su}(2)\;.
\end{align}
For $G$ a simply laced Lie algebra one finds  
\begin{align}
\textrm{Ind}(\rho_{\rm pr}^G )  = \frac{1}6 d_G h_G\;,
\end{align}
where $d_G$ and $h_G$ are the dimension and Coxeter number given in Table~\ref{table:LieA}. This identity can be shown by direct evaluation using the branching rules in \eqref{branching rule}, see \cite{panyushev2009dynkin} for a rigorous proof. The classical CS action for $\mathcal{A}^{\rm geom}_{G=SU(2)} = (\omega + i e)$ gives two basic topological invariants of the hyperbolic 3-manifold, the hyperbolic volume ${\rm vol}(M_3)$ and the Chern-Simons invariant ${\rm cs}(M_3)$, 
\begin{align}
S_0^{\rm geom}[M_3, SU(2)_\mathbb{C}] = i \left( {\rm vol}(M_3)+ i\, {\rm cs} (M_3)\right)\;.
\end{align}
In our convention, the Chern-Simons invariant is defined  modulo $\pi^2$.  
We thus arrive at the following explicit form of the classical CS action for the $\mathcal{A}_G^{\rm geom}$ flat connection
\begin{align}
S_0^{\rm geom}[M_3, G_\mathbb{G}]   =(S_0^{\overline{\rm geom}}[M_3, G_\mathbb{G}] )^*=\frac{i}6\,d_G h_G\left( \textrm{vol}(M_3) + i\,{\rm cs}(M_3)\right)
\;. \label{classical part} 
\end{align}
The 1-loop perturbative invariants $S^\alpha_1$ for $\alpha = (\textrm{geom})$ and  $(\overline{\textrm{geom}}) $ can be written as
\begin{align}
\begin{split}
&S_1^{\rm geom}[M_3,G_\mathbb{C}]  =\left( S_1^{\overline{\rm geom}}[M_3,G_\mathbb{C}]  \right)^*  = -\frac{1}2  \log    \textrm{Tor}_{\rm Adj}[\mathcal{A}^{\rm geom}_G ;M_3, G_{\mathbb{C}}] 
\\
&=  -\frac{1}2  \sum_m \mathcal{N}^{G}_{{\rm Adj}, \tau_{2m+1}} \log \textrm{Tor}_{\rm \tau_{2m+1} }[\mathcal{A}^{\rm geom}_{SU(2)};M_3, SL(2,\mathbb{C})]  \;.
\label{S1}
\end{split}
\end{align}
Here $\mathcal{N}^{G}_{\rm Adj, \tau_{2m+1}} \in \{0, 1\}$ is the number of times the $\tau_{2m+1}$ representation of $SU(2)$ appears in the branching rules \eqref{branching rule}, i.e. we can rewrite \eqref{branching rule} as
\begin{align}
&{\rm Adjoint \;of } \;G  \rightarrow  \bigoplus_{m} \mathcal{N}^G_{\rm Adj, \tau_{2m+1}} \times  \tau_{2 m+1}.
\end{align}
$\textrm{Tor}_{\rm \tau_{2m+1} }[\mathcal{A}^{\rm geom}_{SU(2)};M_3, SL(2,\mathbb{C})] $  is the Ray-Singer-Reidemeister torsion twisted by an $SL(2,\mathbb{C})$ flat connection  $\mathcal{A}^{\rm geom}_{SU(2)}$ in the  representation $\tau_{2m+1}$. The above relation  \eqref{S1} simply follows from the definition in \eqref{Def : Torsion} of the torsion. Using the mathematical results in \cite{park2016reidemeister,park2017reidemeister} one finds the important relation
\begin{align}
\log \big{|}\textrm{Tor}_{\rm \tau_{2m+1} }[\mathcal{A}^{\rm geom}_{SU(2)};M_3, SL(2,\mathbb{C})] \big{|} = \frac{1}{\pi} \textrm{vol}(M_3) \left(m^2+m+\frac{1}6\right) + \sum_{\gamma}\sum_{k=m+1}^{\infty} \log |1-q_\gamma^k|\;. \label{Selberg-formula}
\end{align}
The sum runs over all primitive geodesics $\gamma$ on $M_3$ and we have defined $q_\gamma := e^{-\ell_{\mathbb{C}}(\gamma)}$ where $\ell_{\mathbb{C}}(\gamma)$ is the complexified length
\begin{align}
\textrm{Tr} P \exp \left(\oint_\gamma \mathcal{A}^{\rm geom}_{SU(2)}\right) = e^{\frac{1}2\ell_{\mathbb{C}}(\gamma)} +e^{-\frac{1}2\ell_{\mathbb{C}}(\gamma)}\;, \; \quad \textrm{Re} (\ell_{\mathbb{C}})>0\;.
\end{align}
The real part $\textrm{Re} (\ell_{\mathbb{C}})$ measures the geodesic length. The infinite sum  in \eqref{Selberg-formula} converges absolutely when $m>1$. If the term with $m=1$ does not converge, we cannot use this result.  This is however not important in our context since any potential modification at $m=1$ will affect only the $N^0$ part of the large $N$ free energy of the class $\mathcal{R}$ theories.

 Combining \eqref{branching rule}, \eqref{S1}, and \eqref{Selberg-formula}, we arrive at
\begin{align}
\textrm{Re}(S_1^{\rm geom}[M_3, G_{\mathbb{C}}]) = \textrm{Re}(S_1^{\overline{\rm geom}}[M_3, G_{\mathbb{C}}]) = -\frac{1}{12 \pi } (2d_G h_G + r_G)\textrm{vol}(M_3)-\frac{1}2 R_G(M_3)
\;.
\label{1-loop part}
\end{align}
To obtain this formula we used the following group theoretical fact
\begin{align}
\sum_{m} \mathcal{N}^G_{{\rm Adj},\tau_{2m+1}}\times  \left(m^2+ m+\frac{1}6 \right) = \frac{1}6 \left(2d_G h_G +r_G\right)\;,
\end{align}
which can be shown by using the branching rules in \eqref{branching rule}. The quantity $ R_G(M_3)$ is defined as
\begin{align}
\begin{split}
R_G (M_3) &:= \sum_{m,[\gamma]}  \sum_{k=m+1}^\infty \mathcal{N}^G_{{\rm Adj},\tau_{2m+1}}\times  \log |1-q_\gamma^k| \;,
\\
&= \textrm{Re}   \sum_{m,[\gamma]}  \sum_{k=m+1}^\infty   \mathcal{N}^G_{{\rm Adj},\tau_{2m+1}} \log \textrm{P.E.} \left[-q_\gamma^k \right]\;.
\end{split}
\end{align}
Here ${\rm P.E.}$ is the Plethystic exponential
\begin{align}
\textrm{P.E.} [f(q)]:= \exp \left( \sum_{n=1}^\infty  \frac{1}n f(q^n) \right)\;.
\end{align}
We also note that for $G=A_{N-1}$ and $D_N$ the quantity $R_G(M_3)$ is $O(N^0)$ at large $N$ and takes the explicit form
\begin{align}
\begin{split}
&R_{A_{N-1}}(M_3) = \textrm{Re} \sum_{[\gamma]} \log \textrm{P.E.}\left[\frac{q_\gamma^{N+1}-q_\gamma^2}{(1-q_\gamma)^2}\right]\;,
\\
&R_{D_{N}}(M_3) = \textrm{Re} \sum_{[\gamma]} \log \textrm{P.E.}\left[\frac{(q_\gamma^{N}+q_\gamma^2)(q_\gamma^N-1)}{(1-q_\gamma)^2(1+q_\gamma)}\right]\;.
\end{split}
\end{align}
Note that $q_\gamma^N $ is exponentially small  at large $N$ and thus both $R_{A_{N-1}}(M_3)$ and $R_{D_N}(M_3)$  are $O(N^0)$ at large $N$. 

So far we have discussed the classical and 1-loop contribution to the CS partition function. It is clear from the results above that these two terms contribute to the leading and subleading terms in the large $N$ approximation for the CS partition. It is not known in general how the rest of the perturbative CS invariants, $S_n^{\alpha}$ for $n\geq 2$, behave at large $N$. As pointed out in \cite{Gang:2014ema}, for the $A_N$ CS theory the holographic results for the squashed sphere partition function strongly suggest that $S_2^{\alpha}$ scales as $N^3$ and $S_n^{\alpha}$ are subleading for $n\geq 3$. We provide further evidence for this conjecture below and extend it to the $D_N$ series and to subleading order in the large $N$ expansion.

\subsection{3d-3d relations and the large $N$ limit} 

Here we present explicit 3d-3d relations for various supersymmetric backgrounds $\mathbb{B}$ written in terms of perturbative invariants studied in the previous section. For completeness, we do not suppress the imaginary parts of the free energies~$\log Z$ in this section. Also, for simplicity, we focus on  the case with trivial $H^1(M_3, Z_G)$. There are several subtle issues for the case with non-trivial $H^1(M_3, Z_G)$ \cite{Benini:2019dyp,Eckhard:2019jgg,Cho:2020ljj}. But, the subtleties only affect the $\log N$ and $O(N^0)$ terms in the free energy at large $N$ and thus the large $N$ expansion formula below is  valid for general $M_3$ up to $O(N)$.  

\subsubsection*{$\mathbb{B}= \mathcal{M}_{\mathfrak{g},p} $ (degree $p$ bundle over $\Sigma_{\mathfrak{g}}$) }

The 3d-3d relation for this supersymmetric background is \cite{Gang:2018hjd,Gang:2019uay}
\begin{align}
\begin{split}
&Z\left[T[M_3,G] \textrm{ on } \mathcal{M}_{\mathfrak{g},p\in 2\mathbb{Z}} \right] 
\\
&=  \sum_{\mathcal{A}^\alpha \in \chi_{\rm irred}\left(M_3, G_{\mathbb{C}}\right)} \left( \frac{1}{|Z_G|}  \exp ( 2 S_1^\alpha [M_3, G_{\mathbb{C}}]  \right)^{1-\mathfrak{g}} \exp \left( - p \frac{S_0^\alpha[M_3, G_{\mathbb{C}}]}{2\pi i } \right)\;. \label{3d-3d relation I}
\end{split}
\end{align}
Note that the above 3d-3d relation  works only for even $p$. For even $p$, there are two supersymmetric backgrounds depending on  spin-structure choices along the fiber $S^1$-direction~\cite{Closset:2018ghr}. The 3d-3d relation is for the anti-periodic boundary condition along the fiber $S^1$-direction.  For $p=0$, the partition function computes the following twisted index
\begin{align}\label{eq:TTI}
Z\left[T[M_3,G] \textrm{ on } \mathcal{M}_{\mathfrak{g},p=0} \right]  = \textrm{Tr}_{\mathcal{H}_{\rm top}(\Sigma_g)}(-1)^R\;.
\end{align}
Here $\mathcal{H}_{\rm top}(\Sigma_\mathfrak{g})$ is the Hilbert-space of the $T[M_3, G]$ theory on a Riemann surface $\Sigma_\mathfrak{g}$ with a topological twisting using the $U(1)$ R-symmetry. With $R$ in \eqref{eq:TTI} we denote the charge with respect to the R-symmetry. The $U(1)$ R-symmetry originates from the $SO(2)$ subgroup of the $SO(5)$ R-symmetry of 6d $(2,0)$ theory which commutes with the $SO(3)$ subgroup used for the class $\mathcal{R}$ topological twist. This in turn implies that the R-charge is integer valued for all states in the Hilbert-space.  The integrality of the R-charge guarantees that the Dirac quantization condition, $(\mathfrak{g}-1)R \in \mathbb{Z}$, is obeyed.  At sufficiently large $N$, there is no continuous flavor symmetry in the class $\mathcal{R}$ theory and the compact $U(1)$ R-symmetry is the IR superconformal R-symmetry. This situation should be contrasted with 3d $\mathcal{N}=2$ SCFTs arising as world-volume theory on M2 branes probing a conical CY 4-fold over a 7d Sasaki-Einstein manifold. In those examples the Dirac quantization generically can not be satisfied  for the superconformal R-symmetry  and thus the supersymmetric partition on $\mathcal{M}_{\mathfrak{g},p}$ is ill-defined \cite{Toldo:2017qsh}. 
As a consistency check of the 3d-3d relation in \eqref{3d-3d relation I}, one can confirm that the right hand side becomes an integer when $p=0$ \cite{Gang:2019uay}. 

For $p>0$ (resp. $p<0$), the flat connection $\mathcal{A}_G^{\alpha = (\rm geom)}$ (resp. $\mathcal{A}_G^{\alpha = (\overline{\rm geom})}$) gives the dominant contribution at large $N$. By combining \eqref{classical part} and \eqref{1-loop part} with \eqref{3d-3d relation I}, we find 
\begin{align}
\begin{split}
&\log \big{|}Z\left[T[M_3,G= A_{N-1} \textrm{ or }D_N] \textrm{ on } \mathcal{M}_{\mathfrak{g}>1,p_{\in 2\mathbb{Z}}>0} \right] \big{|}
\\
& \xrightarrow{\quad N\rightarrow \infty\quad  }   \frac{\textrm{vol}(M_3)}{6\pi } \left((\mathfrak{g}-1)(2d_G h_G +r_G)+\frac{p}2 d_G h_G \right) + (\mathfrak{g}-1) \log |Z_G| +(\mathfrak{g}-1) R_G (M_3) 
\\
& \qquad \qquad \quad + O(N^0)\;.
\end{split}
\end{align}
Recall that $Z_G$ is the center group of $G$ and $|Z_G|$ is its order.  For $p=0$, on the other hand, the dominant contribution at large $N$ is given by two flat connections,  $\mathcal{A}_G^{\alpha = (\rm geom)}$ and  $\mathcal{A}_G^{\alpha = (\overline{\rm geom})}$,  and one finds: 
\begin{align}\label{eq:logZp0largeN}
\begin{split}
&\log \big{|} Z\left[T[M_3,G= A_{N-1} \textrm{ or }D_N] \textrm{ on } \mathcal{M}_{\mathfrak{g}>1,p=0} \right] \big{|} 
\\
& \xrightarrow{\quad N\rightarrow \infty \quad }   \frac{\textrm{vol}(M_3)}{6\pi } (\mathfrak{g}-1)(2d_G h_G +r_G) + (\mathfrak{g}-1) \log |Z_G| +(\mathfrak{g}-1) R_G (M_3) 
\\
& \qquad \qquad \quad + (\mathfrak{g}-1) \log \left(2\cos (\theta_{M_3, G})\right) + O(N^0)\;.
\end{split}
\end{align}
The  term $(\mathfrak{g}-1) \log \left(2\cos (\theta_{M_3, G})\right)$ for $p=0$ comes form the relative phase factor between the two dominant flat connections
\begin{align}
\begin{split}
&\textrm{Tor}_{\rm Adj} [\mathcal{A}^{\rm geom}_G;M_3, G_\mathbb{C}]= e^{i \theta_{M_3, G}}  \big{|}\textrm{Tor}_{\rm Adj} [\mathcal{A}^{\rm geom}_G;M_3, G_\mathbb{C}]\big{|}\;,
\\
&\textrm{Tor}_{\rm Adj} [\mathcal{A}^{\overline{\rm geom}}_G;M_3, G_\mathbb{C}]= e^{-i \theta_{M_3, G}}  \big{|}\textrm{Tor}_{\rm Adj} [\mathcal{A}^{\rm geom}_G;M_3, G_\mathbb{C}]\big{|}\;.
\end{split}
\end{align} 
This large $N$ behavior is the result we have used to fix the constants $(A,B,C)$ in \eqref{eq:tshirtQFT} and arrive at the expression in \eqref{eq:largeN}.  The logarithm term, $(\mathfrak{g}-1)\log N$, in \eqref{eq:logZp0largeN} for $G= A_{N-1}$ can be reproduced holographically by a 1-loop supergravity computation \cite{Gang:2018hjd}. For $G=D_N$, on the other hand, there is no $\log N$ term since then $|Z_G|=4=O(N^0)$. It would be interesting to understand the absence of a $\log N$ term for $G=D_N$ from the supergravity side.  

\subsubsection*{$\mathbb{B}= S^1\times_{\omega}S^2$ (superconformal index)}

To compute the superconformal index for theories of class $\mathcal{R}$ using the 3d-3d correspondence one has to take the limit $\omega \to 0$ which is akin to the Cardy limit familiar from 2d CFTs. For hyperbolic $M_3$, the 3d-3d relation in the Cardy limit reads \cite{Dimofte:2011py}
\begin{align}
\begin{split}
&Z\left[T[M_3,G] \textrm{ on } S^2 \times_q S^1\right] \xrightarrow{\quad q=e^{2\pi i \omega} , \;\omega \rightarrow 0 \quad } 
\\
&\frac{1}{|Z_G|}  \sum_{\mathcal{A}^\alpha \in \chi_{\rm irred}\left(M_3, G_{\mathbb{C}}\right)}\exp \left( \sum_{n=0}^{n=\infty} \big{(}\hbar^{n-1} S^{\alpha}_n [M_3, G_{\mathbb{C}} ] +(-\hbar)^{n-1} S^{\overline{\alpha}}_n  [M_3, G_{\mathbb{C}}] \big{)} \right) \bigg{|}_{\hbar = 2\pi i \omega} \;. \label{3d-3d relation II}
\end{split}
\end{align}
Here $S^{\overline{\alpha}}_n = (S^\alpha_n)^*$ is the perturbative expansion coefficient associated to the complex conjugate flat connection $\mathcal{A}^{\overline{\alpha}} = (\mathcal{A}^\alpha)^*$.  This partition function is equivalent to the following superconformal index
\begin{align}
\textrm{Tr}_{\mathcal{H}_{\rm rad}(S^2)} (-1)^R q^{\frac{R}2 + j_3}\;.
\end{align}
The trace is taken over the Hilbert space in radial quantization, whose elements are local operators in the $T[M_3, G]$ theory. In the Cardy limit with $\textrm{Re}(\omega)<0$, from the property in \eqref{property-of-A-geom}, the flat connection $\mathcal{A}^{\overline{\rm geom}}_G$ gives the most dominant contribution. Using  \eqref{classical part} and \eqref{1-loop part}, we find
\begin{equation}
\begin{split}
&\log \big{|} Z\left[T[M_3,G] \textrm{ on } S^2 \times_q S^1\right] \big{|} \xrightarrow{\quad q=e^{2\pi i \omega}, \;\omega \rightarrow 0^- \quad } 
\\
& -  \frac{{\rm vol}(M_3)}{6\pi } \left(\frac{d_G h_G}\omega + 2d_G h_G+r_G \right)  -\log |Z_G|    - \frac{R_G (M_3)}{6\pi}  
\\ 
& +2\sum_{n=1}^\infty (-1)^n \textrm{Im}[S^{\overline{\rm geom}}_{2n}[M_3,G_\mathbb{C}] ](2\pi  \omega)^{2n-1} +2\sum_{n=1}^\infty (-1)^n \textrm{Re}[S^{\overline{\rm geom}}_{2n+1}[M_3,G_\mathbb{C}] ](2\pi  \omega)^{2n}    \;. 
\end{split}
\end{equation}
This result is compatible with the $\omega \to 0$ limit of the holographic prediction in \eqref{eq:largeN} up to $O(\omega^0)$. By comparing the higher order terms in $\omega$ above with the holographic prediction,  we arrive at the following non-trivial mathematical predictions
\begin{align}
\begin{split}
& \textrm{Im}[S^{\overline{\rm geom}}_2[M_3, G] ] = \frac{1}{24\pi^2}\textrm{vol}(M_3)\,d_G h_G +O(N^0)\;,
 \\
& \textrm{Im}[S^{\overline{\rm geom}}_{2n}[M_3, G] ]_{n>1} = O(N^0)\;, \quad  \textrm{Re}[S^{\overline{\rm geom}}_{2n+1}[M_3, G] ]_{n\geq 1} = O(N^0)
\end{split}
   \label{two-loop}
\end{align}
A similar conjecture was already proposed in \cite{Gang:2014ema} for $G=A_{N-1}$ case. However, in \cite{Gang:2014ema} the conjecture was only about  the leading order in the large $N$ limit, i.e. the $O(N^3)$ term in \eqref{two-loop}. The HD holographic results summarized in Section~\ref{sec:HoloFBH} allow us to generalize this up to $O(N^1)$ as well as to the $D_N$ class of theories. It would be very interesting to confirm the conjecture in \eqref{two-loop} by other means. For example by direct numerical evaluation of $S_{n\geq 2}^{\overline{\rm geom}}[M_3, G=A_{N-1}]$ for various hyperbolic 3-manifold $M_3$ and different values of  $N$.  These calculations should be facilitated by the techniques developed in \cite{Dimofte:2012qj,Dimofte:2013iv,Gang:2017cwq}.

\subsubsection*{$\mathbb{B}= S^3_b$ (squashed $S^3$ partition function)}

To calculated this partition function using the 3d-3d correspondence we again have to appeal to a specific limit of the squashing parameter resembling a Cardy-like limit, namely $b \to 0$. For hyperbolic $M_3$, the 3d-3d relation in this Cardy-like limit is \cite{Gang:2014ema} 
\begin{align}
Z\left[T[M_3,G] \textrm{ on } S^3_b\right] \xrightarrow{\quad b^2 \in \mathbb{R}_+ \rightarrow 0 \quad } \sqrt{\frac{1}{|Z_G|} } \exp \left( \sum_{n=0}^{n=\infty} \hbar^{n-1} S^{\overline{\rm geom}}_n  [M_3,  G_{\mathbb{C}}]  \right) \bigg{|}_{\hbar = 2\pi i b^2}  \;. \label{3d-3d relation III}
\end{align}
Note that only a single  irreducible flat connection, $\mathcal{A}^{\overline{\rm geom}}$, contributes in the Cardy limit, see \cite{Gang:2014ema,Bae:2016jpi,Gang:2017hbs,Mikhaylov:2017ngi}. Using the formulae in \eqref{classical part}, \eqref{1-loop part}, and \eqref{two-loop}, we find the following result 
\begin{align}
\begin{split}
&\log \big{|}Z\left[T[M_3,G] \textrm{ on } S^3_b\right] \big{|} \xrightarrow{\quad b^2 \in \mathbb{R}_+ \rightarrow 0 \quad }
\\
&  -\frac{\textrm{vol}(M_3)}{12\pi }\left(  \frac{d_G h_G }{b^2}+2d_G h_G +r_G +d_G h_G b^2 \right)- \frac{1}2 \log |Z_G| -\frac{R_G(M_3)}{12\pi} + O(N^0)  \;. 
\end{split}
\end{align} 
This expression is compatible with the $b\to 0$ limit of the holographic prediction in \eqref{eq:largeN}. This provides additional non-trivial evidence for the consistency of all results presented above and for the validity of the conjecture in \eqref{two-loop}. 
%
\section{Discussion}
\label{sec:discussion}

In this paper we combined results from four-derivative 4d $\mathcal{N}=2$ gauged supergravity, holography, and the 3d-3d correspondence to shed light on the physics of the 3d $\mathcal{N}=2$ theories of class $\mathcal{R}$ in the large $N$ approximation. Our results amount to explicit evaluation of the leading and subleading terms in the large $N$ expansion of various supersymmetric partition functions and the two-point function of the energy momentum tensor. In addition we are able to calculate explicitly the first subleading correction to the Bekenstein-Hawking entropy for any asymptotically AdS$_4$ black hole solution arising from M5-branes wrapped on a hyperbolic manifold.

Our work leads to a number of open questions and avenues for generalization. Here we briefly discuss a few of them.

\begin{itemize}

\item We have studied the leading correction to the $N^3$ behavior of a number of physical observables in the large $N$ limit of the class $\mathcal{R}$ 3d $\mathcal{N}=2$ QFTs. Studying the behavior of more subleading corrections is very interesting but challenging both on the QFT and gravity side. In supergravity one needs to study corrections to the supergravity action that involve six or more derivatives while in the 3d-3d correspondence we need detailed knowledge about the large $N$ behavior of the perturbative CS invariants $S_n^{\alpha}$ for $n\geq2$. Progress on both of these fronts will be very interesting. 

\item As described in the introduction we have studied higher derivative corrections to supergravity by using a 4d approach and thus circumvented the need to work with the higher-derivative corrections to 11d supergravity. Understanding how to uplift our explicit results to 11d and map them to coefficients of the supergravity effective action, along the lines of \cite{Chester:2018aca,Binder:2018yvd,Chester:2018dga}, is a very interesting topic for further exploration.

\item Our supergravity and holographic results have lead to the conjecture in \eqref{two-loop} for the large $N$ behavior of the perturbative invariants of CS theory with a complexified ADE gauge group on a hyperbolic manifold. It will be most interesting to confirm this conjecture with explicit calculations along the lines of \cite{Gang:2014ema} or furnish a general proof.

\item In our analysis we have assumed that the hyperbolic manifold $M_3$ is smooth and compact. This was necessitated by the use of the supergravity consistent truncation results and the restriction to work in the minimal 4d $\mathcal{N}=2$ gauged supergravity. It should be possible to generalize this setup by including defects and boundaries on $M_3$ which support additional degrees of freedom in M-theory. In the context of 4d $\mathcal{N}=2$ gauged supergravity these extra degrees of freedom should be incorporated by the addition of vector and hyper multiplets. The probe brane analysis in \cite{Bah:2014dsa} may be useful in uncovering the details of this setup.

\item All QFT results for the 3d $\mathcal{N}=2$ theories of class $\mathcal{R}$ we have used are obtained by using the 3d-3d correspondence to map the calculation to complexified CS theory on the hyperbolic manifold $M_3$. It will be interesting to understand whether some of these quantities can be computed in a more direct manner using the properties of the 3d $\mathcal{N}=2$ theory itself.

\item Here we have focused on twisted compactifications of the $\mathcal{N}=(2,0)$ theory to three dimensions which preserve 3d $\mathcal{N}=2$ supersymmetry. Geometrically these twists can be realized by wrapping M5-branes on special Lagrangian submanifolds in non-compact CY 3-folds. There is a generalization of this construction to twisted compactifications preserving only 3d $\mathcal{N}=1$ supersymmetry which is realized geometrically by M5-branes wrapping associative cycles in non-compact G$_2$ manifolds. Indeed, this generalization has been studied both on the supergravity and field theory side, see \cite{Acharya:2000mu} and \cite{Eckhard:2018raj}, respectively. It will be very interesting to generalize some of our results to this less supersymmetric setup. We expect this to be non-trivial and plagued by technical difficulties due to the small amount of supersymmetry and the lack of explicit results for the large $N$ limit of the $\mathcal{N}=1$ 3d-3d correspondence.

\item It will be interesting to perform a 1-loop supergravity calculation, along the lines of \cite{Gang:2018hjd}, and confirm that the logarithmic tem in \eqref{eq:logZp0largeN} for  $G=D_N$ indeed does not behave as $\log N$.

\item The 3d-3d results for the superconformal index  in \eqref{3d-3d relation II} and the squashed sphere partition function in \eqref{3d-3d relation III} are valid in a Cardy-like limit where the length of an $S^1$ in the geometry is vanishing. On the other hand, the supergravity results in \eqref{eq:largeN} are valid for general values of the squashing parameter $b$ and the fugacity $\omega$. It is desirable to extend the range of applicability of the 3d-3d correspondence and calculate these two partition functions for general values of the parameters.

\item The expression for the partition function in \eqref{eq:logZintro} bears a strong resemblance to the structure of the anomaly polynomial of a 6d $\mathcal{N}=(2,0)$ SCFT of type $G$. It is tempting to speculate that \eqref{eq:logZintro} may be obtained by a suitable equivariant integration of this anomaly polynomial.

\item Wrapped M5-branes lead to a rich family of 4d $\mathcal{N}=2$ and $\mathcal{N}=1$ as well as 2d $\mathcal{N}=(0,2)$ SCFTs which have an explicit holographically dual description, see \cite{Maldacena:2000mw,Gaiotto:2009gz,Bah:2012dg} and \cite{Gauntlett:2000ng,Benini:2013cda}, respectively. It will be very interesting to study these SCFTs using higher-derivative corrections to supergravity. Some concrete results in this spirit were obtained in \cite{Baggio:2014hua} and we hope they could be generalized significantly.

\end{itemize}

\section*{Acknowledgments}
We are grateful to Chris Beem, Francesco Benini, and Marcos Crichigno for valuable discussions. We thank the organizers of the ``QFT and Geometry'' online seminar series for hosting an interactive forum for discussion which provided the impetus for this project.  NB is supported in part by an Odysseus grant G0F9516N from the FWO. The research of DG is supported in part by the National Research Foundation of Korea under grant 2019R1A2C2004880. DG also acknowledges support by the appointment to the JRG program at the APCTP through the Science and Technology Promotion Fund and Lottery Fund of the Korean Government, as well as support by the Korean Local Governments, Gyeongsangbuk-do Province, and Pohang City. KH is supported in part by the Bulgarian NSF grants DN08/3, N28/5, and KP-06-N 38/11. NB, AMC, and VR are also supported by the KU Leuven C1 grant ZKD1118 C16/16/005.

\bibliography{HD-AdS4-classR}
\bibliographystyle{JHEP}

\end{document}